\documentclass[journal=jacsat,manuscript=article]{achemso}

\usepackage{chemformula} 
\usepackage[T1]{fontenc} 
\usepackage[version=4]{mhchem}
\usepackage{adjustbox}
\usepackage[colorlinks=true, allcolors=blue]{hyperref}

\author{Arijit Manna}
\email{arijitmanna@mcconline.org.in}
\phone{+918777749445}
\affiliation{Department of Physics and Astronomy, Midnapore City College, Kuturia, Bhadutala, Paschim Medinipur, West Bengal, India 721129}
\author{Sabyasachi Pal}

\title[Cyanamide towards G31.41+0.31]
  {Detection and chemical modelling of complex prebiotic molecule cyanamide in the hot molecular core G31.41+0.31}


\begin{document}


\begin{abstract}
In the interstellar medium (ISM), the complex prebiotic molecule cyanamide (NH$_{2}$CN) plays a key role in producing adenine (C$_{5}$H$_{5}$N$_{5}$), purines (C$_{5}$H$_{4}$N$_{4}$), pyrimidines (C$_{4}$H$_{4}$N$_{2}$), and other biomolecules via a series of reactions. Therefore, studying the emission lines of NH$_{2}$CN is important for understanding the hypothesis of the pre-solar origin of life in the universe. We present the detection of the rotational emission lines of NH$_{2}$CN with vibrational states $v$ = 0 and 1 towards the hot molecular core G31.41+0.31 using the high-resolution twelve-meter array of Atacama Large Millimeter/Submillimeter Array (ALMA) band 3. The estimated column density of NH$_{2}$CN towards G31.41+0.31 using the local thermodynamic equilibrium (LTE) model is (7.21$\pm$0.25)$\times$10$^{15}$ cm$^{-2}$ with an excitation temperature of 250$\pm$25 K. The abundance of NH$_{2}$CN with respect to H$_{2}$ towards G31.41+0.31 is (7.21$\pm$1.46)$\times$10$^{-10}$. The NH$_{2}$CN and NH$_{2}$CHO column density ratio towards G31.41+0.31 is 0.13$\pm$0.02. We compare the estimated abundance of NH$_{2}$CN with that of other hot cores and corinos and observed that the abundance of NH$_{2}$CN towards G31.41+0.31 is nearly similar to that of the hot molecular core G358.93--0.03 MM1, the hot corinos IRAS 16293-2422 B, and NGC 1333 IRAS4A2. We compute the two-phase warm-up chemical model of NH$_{2}$CN using the gas-grain chemical code UCLCHEM, and after chemical modelling, we notice that the observed and modelled abundances are nearly similar. After chemical modelling, we conclude that the neutral-neutral reaction between NH$_{2}$ and CN is responsible for the production of NH$_{2}$CN on the grain surface of G31.41+0.31.
\end{abstract}
\text{keywords:}\\
\text{{\small astrochemistry, complex organic molecules, hot molecular cores, high-mass star formation region, ISM}}

\section{1. Introduction}
\label{sec:intro} 
The complex prebiotic molecule NH$_{2}$CN contains both cyan and amide groups, and it plays an important role in the study of the origin of life in the universe \citep{st64, br99}. NH$_{2}$CN is a possible precursor of urea (NH$_{2}$CONH$_{2}$) and is known to be a key compound in the investigation of life in star-forming regions \citep{lig20}. One of the less stable isomers of NH$_2$CN in the interstellar medium (ISM) is carbodiimide (HNCNH) \citep{tur75, jab19}. The HNCNH molecule contains an --NCN-- frame, which plays a major role in assembling amino acids to the peptides in liquid water (H$_{2}$O) \citep{wil81}. NH$_{2}$CN contains 0$^{-}$ and 0$^{+}$ substates \citep{mil62}. The magnetic dipole moments of the 0$^{-}$ and 0$^{+}$ substates are 4.24 debye and 4.25 debye, respectively \citep{bro85}. The emission lines of NH$_{2}$CN were detected in the hot molecular cores Sgr B2 (N) \citep{tur75, bel13}, Sgr B2 (M) \citep{bel13}, IRAS 20126+410 \citep{whi03}, Orion KL \citep{pa17}, G10.47+0.03 \citep{man22}, G358.93--0.03 MM1 \citep{man23a}, solar-like protostars IRAS 16293--2422 B \citep{cou18} and NGC 1333 IRAS2A \citep{cou18}, starburst galaxy M82 \citep{mar06}, and sculptor galaxy NGC 253 \citep{ala11}.

The study of complex molecular lines in star-forming regions using (sub)millimeter-wavelength telescopes is important for understanding the prebiotic chemistry in these regions. Hot molecular cores are early phases of massive star-formation regions, and most complex molecules, including glycine (NH$_{2}$CH$_{2}$COOH) precursors (\ce{NH2CH2CN}, \ce{CH2NH}, \ce{CH3NH2}, etc.), are frequently found in hot molecular cores \citep{her09, shi21, man22a}. The complex molecules in the hot molecular cores significantly increase the chemical richness of the ISM. Hot molecular cores are suitable candidates for studying individual complex organic molecules because they have a high gas density ($\geq$ 10$^{6}$ cm$^{-3}$), a small and compact source size ($\leq$ 0.1 pc), and a warm gas temperature ($\geq$ 100 K) \citep{van98, vi04}. Most of the complex molecular lines arise mainly from the warm inner region of the envelope of hot cores \citep{van98, her09, man231}. The abundance of complex molecules varies from $\sim$10$^{-11}$ to $\sim$10$^{-7}$ \citep{her09}. Most complex molecules are formed on the icy surface of hot molecular cores by hydrogenation of simple icy species \citep{gar06, gar08, gar13}. In hot molecular cores, the gas temperature rapidly increases because of the collapse of the prestellar core, which is called the ``warm-up" phase \citep{gar13}. The maximum number of complex organic molecules in the hot molecular cores is created and destroyed during the warm-up phase \citep{gar13}. The formation pathways of complex organic molecules in hot molecular cores have been intensively debated in astrochemistry. In hot cores and corinos, two possible pathways have been proposed to produce complex organic molecules: grain surface and gas-phase chemical reactions \citep{vi04, gar06, gar08, gar13}. A comparison between the two-phase \citep{vi04} and three-phase \citep{gar13} warm-up chemical modelling abundances and the observed abundances is necessary to understand the possible formation pathways of complex organic molecules in hot molecular cores.

\begin{table*}{}
	\centering
	\caption{Observation summary of hot molecular core G31.41+0.31.}
	\begin{adjustbox}{width=1.0\textwidth}
		\begin{tabular}{cccccccccccc}
			\hline
Date of observation  &	On-source time&Range of frequency &Beam size    &Spectral resolution &Sensitivity (10 km s$^{-1}$)\\	
(dd-mm-yyyy)     &	(Min)         &(GHz)             &               &(kHz)                 &(mJy beam$^{-1}$)\\
			\hline		
14-03-2018       &	41.328      &98.50 -- 99.44&1.30$^{\prime\prime}$$\times$1.01$^{\prime\prime}$&564.45&	0.55 	\\

--              &   --      &99.41--100.34 &1.30$^{\prime\prime}$$\times$1.02$^{\prime\prime}$  &564.45  &0.55\\

--              &   --           &100.31--101.25&1.29$^{\prime\prime}$$\times$1.01$^{\prime\prime}$  &564.45   &0.55\\

			--               &   --           &101.22--102.15 &1.28$^{\prime\prime}$$\times$1.01$^{\prime\prime}$ &564.45    &0.55\\
			\hline
		\end{tabular}	
	\end{adjustbox}
	\label{tab:data}
\end{table*} 

The disk-like hot molecular core G31.41+0.31 is located at a distance of 7.9 kpc from Earth \citep{ch90}. The luminosity and mass of G31.41+0.31 are 3$\times$10$^{5}$ \textup{L}$_{\odot}$ and 120 \textup{M}$_{\odot}$ \citep{cer10, bel18}. Earlier, \citet{bel18} observed G31.41+0.31 using ALMA at a wavelength of 1.4 mm, and they found two cores around G31.41+0.31, which were defined as the main core and the northeast (NE) core. \citet{bel18} also observed the rotational spin-up, accelerating infall, possible outflow directions, and redshifted absorption towards G31.41+0.31. The ultra-compact (UC) HII region is located on the northeast side of G31.41+0.31, which is separated by $\sim$5$^{\prime\prime}$ from the main core \citep{cer10}. Earlier, the rotational emission lines of several complex molecules were identified towards G31.41+0.31. The rotational emission lines of glycolaldehyde(CH$_{2}$OHCHO), methyl cyanide (CH$_{3}$CN), ethyl cyanide (C$_{2}$H$_{5}$CN), dimethyl ether (CH$_{3}$OCH$_{3}$), ethylene glycol ((CH$_{2}$OH)$_{2}$), methyl formate (CH$_{3}$OCHO), ethanol (C$_{2}$H$_{5}$OH), and other organic molecules have been identified in G31.41+0.31 \citep{iso13, riv17, min20}. Recently, \citet{min23} detected nine oxygen-bearing molecules such as \ce{C2H5OH}, \ce{CH3COCH3}, \ce{CH3OCH3}, \ce{CH3CHO}, aGg$^{\prime}$-(\ce{CH2OH})$_{2}$, gGg$^{\prime}$-(\ce{CH2OH})$_{2}$, $^{13}$\ce{CH3OH}, \ce{CH3OH}, and \ce{CH3}$^{18}$OH towards G31.41+0.31. Similarly, \citet{min23} also detected six nitrogen-bearing molecules such as \ce{C2H5CN}, \ce{C2H3CN}, \ce{C2H5}$^{13}$CN, \ce{CH3CN}, \ce{CH3}$^{13}$CN, and $^{13}$C\ce{H3CN} towards G31.41+0.31 but the authors do not focus to study of the emission lines of \ce{NH2CN}. Recently, the rotational emission line of phosphorus nitride (PN) with transition J = 3--2 was detected towards G31.41+0.31 \citep{man23b}.

\begin{figure*}
	\centering
	\includegraphics[width=0.8\textwidth]{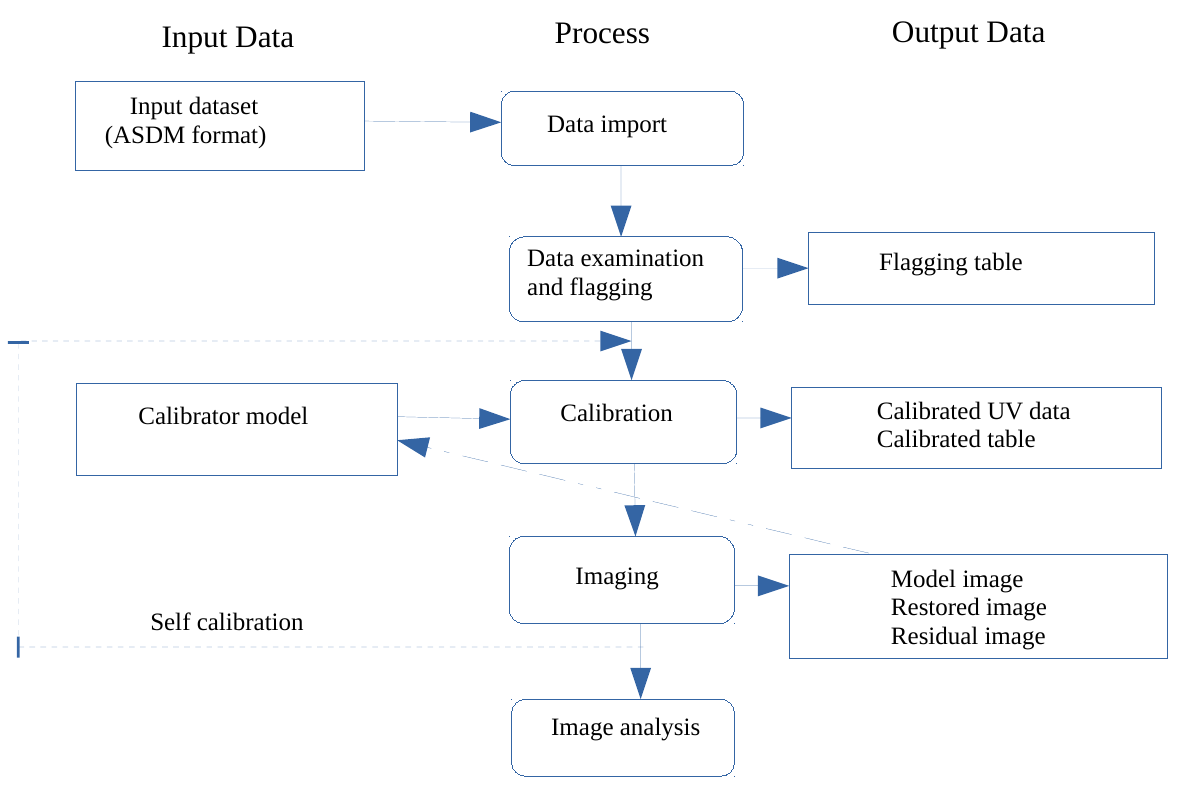}
	\caption{ALMA data analysis flowchart using the CASA. The programme is built on the CASA core, which integrates C++ tools with an interactive Python data reduction interface.}
	\label{fig:flowchart}
\end{figure*}

In this article, we present the identification of the emission lines of the complex nitrogen-bearing molecule NH$_{2}$CN towards G31.41+0.31, using the ALMA band 3. The ALMA observations and data reduction are presented in Section 2. The results of the detection of the emission lines of NH$_{2}$CN and the spatial distribution of NH$_{2}$CN towards G31.41+0.31 are presented in Section~3. The discussion and conclusion of the identification of NH$_{2}$CN in G31.41+0.31 are presented in Sections 4 and 5.

\begin{figure*}
	\centering
	\includegraphics[width=1.0\textwidth]{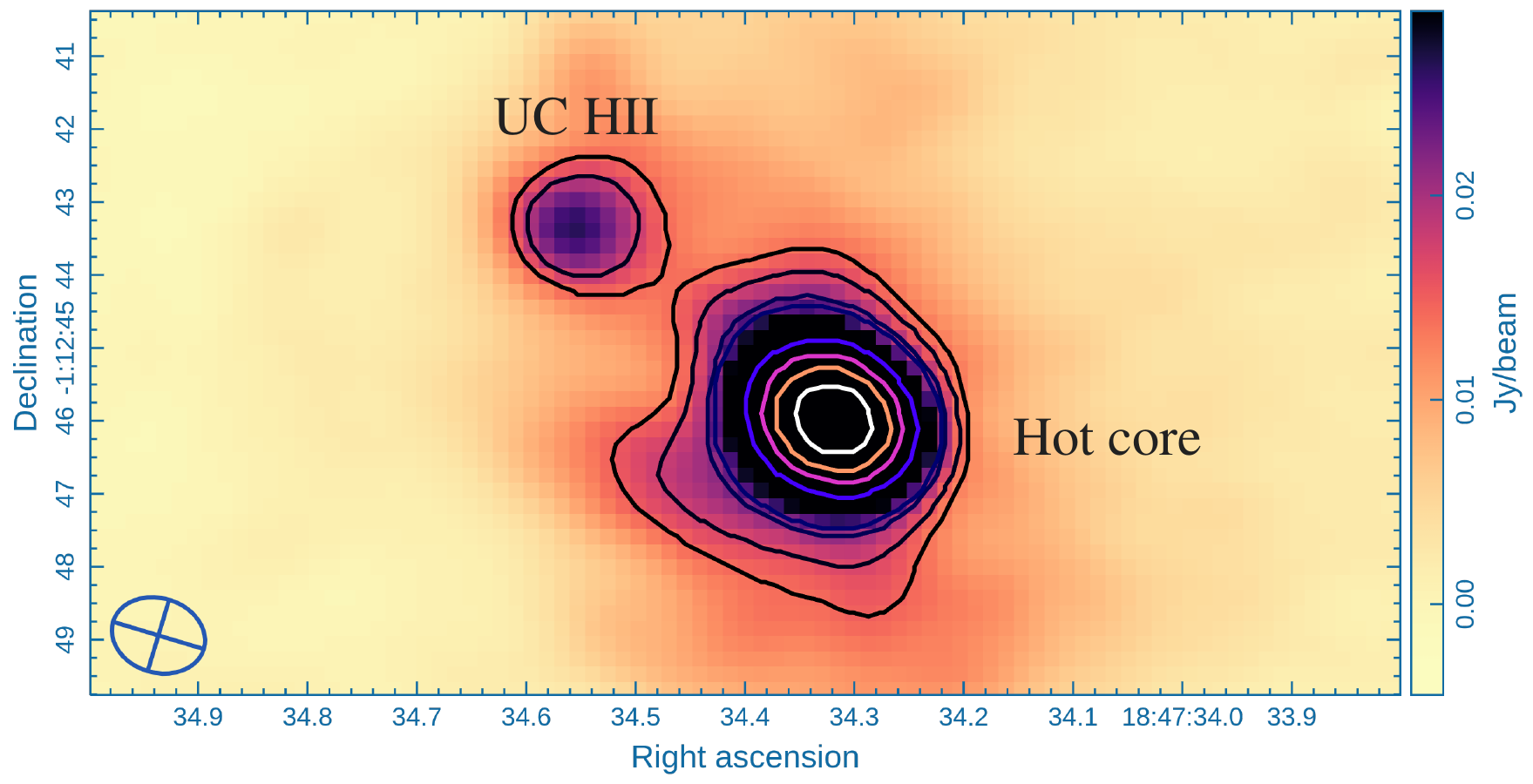}
	\caption{Continuum emission image of G31.41+0.31 at a frequency 99.915 GHz (3.02 mm). The blue circle represents the synthesized beam size of 1.30$^{\prime\prime}\times$ 1.02$^{\prime\prime}$. The contour levels begin at 3$\sigma$ ($\sigma =3.36$ mJy, the root mean square of the background of the continuum emission image) and increase by the factor of $\surd$2.}
	\label{fig:continuum}
\end{figure*}

\section{2. Observations and data reductions}
\label{obs}
We used the archival data of G31.41+0.31, which was observed in the frequency ranges of 84.05 GHz to 115.91 GHz using ALMA band 3 (PI: Beltr{\'a}n Maite). This observation data belongs to the G31.41+0.31 Unbiased ALMA sPectral Observational Survey (GUAPOS)\footnote{\url{https://www.arcetri.inaf.it/~guapos/project.html}} project with ALMA 12m arrays. The phase center of G31.41+0.31 is ($\alpha,\delta$)$_{\rm J2000}$ = 18:47:34.315, --01:12:45.900. This observation was performed to study heavy complex organic molecules, especially those with $\geq$10 atoms, nitrogen-bearing complex organic molecules, such as amino acetonitrile (\ce{NH2CH2CN}), deuterated species, phosphorus, and sulphur-bearing species. The full observations were separated into nine datasets, with each dataset containing four spectral windows. We take only one dataset between the frequency ranges of 98.50 GHz and 102.15 GHz because the higher intensity transitions of NH$_{2}$CN are observed between the frequency ranges of 99.41 GHz and 100.34 GHz in ALMA band 3, which is confirmed by the online molecular database Splatalogue.
	
The ALMA observation of hot molecular core G10.47+0.03 was carried out on March 14, 2018, with an on-source integration time of 41.328 min. A total of forty-three antennas were used, with a minimum baseline of 15.1 m and a maximum baseline of 783.5 m. During the observation, the flux calibrator was taken as J1751+0939, and the phase calibrator was taken as J1851+0035. The observation summary is presented in Table ~\ref{tab:data}.

For data analysis and imaging, we used the Common Astronomy Software Application ({\tt CASA 5.4.1}) with an ALMA data reduction pipeline \citep{mc07}. At first, we used the CASA task {\tt SETJY} with the Perley-Butler 2017 flux calibration model for flux calibration for each baseline \citep{pal17}. Then we used the CASA pipeline tasks {\tt hifa\_bandpassflag} and {\tt hifa\_flagdata} for bandpass calibration and flagging bad antenna data, respectively. After the initial data reduction, we used task {\tt MSTRANSFORM} to split the target source G31.41+0.31 from the rest of the data. After that, we also used task {\tt UVCONTSUB} in the UV plane of the calibrated data to subtract continuum emission from the UV plane. Now the dust continuum images were created from the line-free channels using task {\tt TCLEAN} with {\tt HOGBOM} deconvolution. The molecular spectral line images were created using the task {\tt TCLEAN} with the {\tt SPECMODE=CUBE} parameter. We also performed the repeated self-calibration method using the tasks {\tt GAINCAL} and {\tt APPLYCAL} in an attempt to improve the RMS of the final images. Finally, we used task {\tt IMPBCOR} to correct the synthesized beam pattern in the continuum and spectral images. The data analysis flowchart is shown in Figure~\ref{fig:flowchart}.

\begin{figure*}
	\centering
	\includegraphics[width=1.0\textwidth]{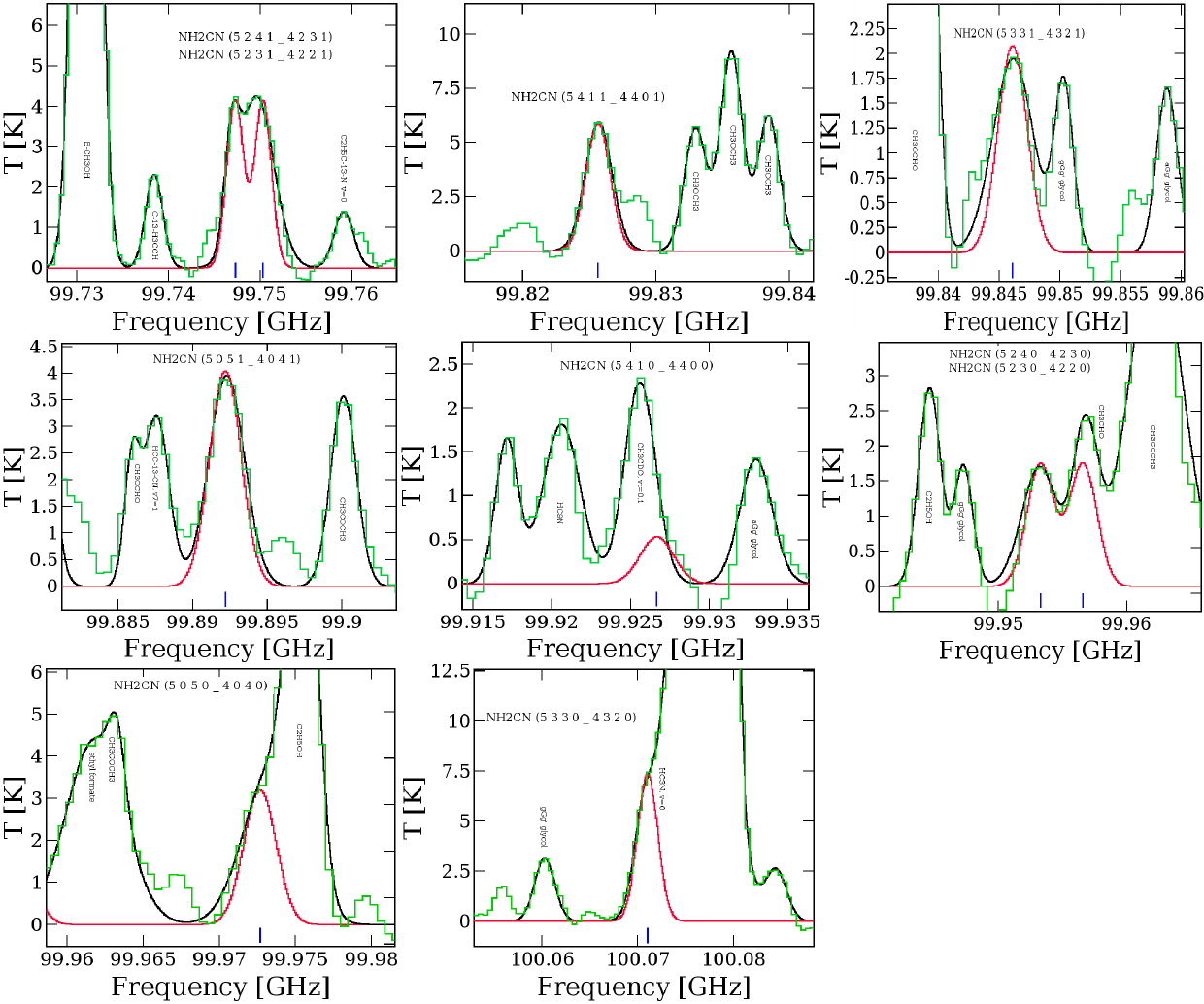}
	\caption{Rotational emission lines of NH$_{2}$CN in G31.41+0.31 with vibrational states $v$ = 0 and 1. The green lines denote the observed molecular spectra of G31.41+0.31. The red lines denote the LTE model spectra of NH$_{2}$CN and the black lines denote the LTE model spectra of other molecules. The vertical blue lines show the positions of the rest frequency of NH$_{2}$CN transitions.}
	\label{fig:emissionspectra}
\end{figure*}

\section{3. Result}
\label{res}
\subsection{3.1 Continuum emission towards the G31.41+0.31}
We present a continuum emission image of the hot molecular core G31.41+0.31 at a frequency of 99.915 GHz, which is shown in Figure~\ref{fig:continuum}. In the continuum emission image, we found that the ultra-compact (UC) HII region is located at a distance of $\sim$5$^{\prime\prime}$ from the main hot core. We fit the two individual 2D Gaussian using the CASA task {\tt IMFIT} over the hot core and UC HII region to estimate the physical properties. The peak and integrated flux densities of the main hot core are 174.4$\pm$3.6 mJy beam$^{-1}$ and 249.3$\pm$7.9 mJy with RMS of 3.36 mJy, respectively. The position angle of the main hot core is 40.27$^{\circ}$. Similarly, the peak and integrated flux densities of the UC HII region are 21.4$\pm$1.3 mJy beam$^{-1}$ and 56.9$\pm$4.8 mJy with RMS of 1.26 mJy, respectively. The position angle of the UC HII region is 44.28$^{\circ}$. The synthesized beam size of the continuum image is 1.30$^{\prime\prime}\times$ 1.02$^{\prime\prime}$. The deconvolved source sizes of the main hot core and UC HII region are 1.56$^{\prime\prime}$$\times$1.46$^{\prime\prime}$ and 1.44$^{\prime\prime}$$\times$1.34$^{\prime\prime}$, respectively. We notice that the deconvolved source sizes of the main hot core and UC HII region are larger than the synthesized beam size of the continuum image. This means the main hot core and UC HII region are resolved at a wavelength of 3.02 mm. Previously, \citet{gor21} reported the detection of continuum emission from the G31.41+0.31 using the ALMA at a frequency of 94.3 GHz with peak flux density 158 mJy beam$^{-1}$ and deconvolved source size 0.90$^{\prime\prime}$$\times$0.68$^{\prime\prime}$.

\subsection{3.2 Detection of the emission lines of NH$_{2}$CN towards G31.41+0.31}
At first, we extracted the molecular spectra from the spectral images of G31.41+0.31 by drawing a 3.1$^{\prime\prime}$ diameter circular region over the main hot core. We could not find any line emission from the UC HII region of G31.41+0.31. The systematic velocity of the spectra of G31.41+0.31 is 97 km s$^{-1}$ \citep{cer10, riv17}. We used the local thermodynamic equilibrium (LTE) model with the Jet Population Laboratory (JPL)\citep{pic98} molecular line list to search the rotational emission lines of NH$_{2}$CN in the spectra of G31.41+0.31. We used CASSIS for the LTE modelling \citep{vas15}. The LTE assumptions are appropriate for G31.41+0.31 because the gas density in the warm inner envelope of G31.41+0.31 is $\sim$1$\times$10$^{8}$ cm$^{-3}$ \citep{cer10, min20}. We applied the LTE-RADEX module in CASSIS to fit the LTE model spectra of NH$_{2}$CN to the observed spectra of the G31.41+0.31. During LTE modelling, we considered a cosmic background temperature of $T_{bg}$ = 2.73 K and background continuum source temperature of $T_{c}$ = 0. Using the LTE model spectra, we have detected a total of ten transition lines of NH$_{2}$CN with vibrational states $v$ = 0 and 1 between the frequency range of 98.50 GHz and 102.15 GHz. After spectral analysis using the LTE model, we find only six transition lines of NH$_{2}$CN that are non-blended, which are detected above 5$\sigma$ significance. We also observe that the transition lines of NH$_{2}$CN at frequencies of 99.926 GHz, 99.956 GHz, 99.972 GHz, and 100.071 GHz are blended with \ce{CH3CDO}, \ce{CH3CHO}, \ce{C2H5OH}, and \ce{HC3N}, respectively. The blended molecules are confirmed by JPL and the online molecular database Splatalogue. The upper-level energies of the detected emission lines of NH$_{2}$CN between the frequency ranges of 98.50 GHz and 102.15 GHz vary between 14.39 K and 311.46 K. There are no missing high-intensity transitions of NH$_{2}$CN in the molecular spectra of G31.41+0.31. The LTE-fitted rotational emission lines of NH$_{2}$CN are shown in Figure~\ref{fig:emissionspectra}. In the spectra, the vertical blue lines indicate the rest frequency positions of the detected \ce{NH2CN} transitions. The LTE-fitted spectral line parameters of NH$_{2}$CN are shown in Table~\ref{tab:MOLECULAR DATA}. Among the ten transition lines of \ce{NH2CN} between the frequency ranges of 98.50 GHz and 102.15 GHz, six non-blended lines are properly fitted with the LTE model spectra of \ce{NH2CN}, and four lines are not properly fitted due to being blended with other molecular transitions. Using LTE modelling, the best-fit column density of NH$_{2}$CN towards G31.41+0.31 is (7.21$\pm$0.25)$\times$10$^{15}$ cm$^{-2}$ with an excitation temperature of 250$\pm$35 K. The source of errors in column density and excitation temperature is the error in integrated intensities ($\rm{\int T_{mb}dV}$) of the emission lines of \ce{NH2CN}, which is obtained by using the least-squares method in the LTE-RADEX module \citep{vas15}. During LTE modelling, we used a source size of 1.3$^{\prime\prime}$. To study the observed emission lines of \ce{NH2CN}, we adopt the hypothetical source size of 1.3$^{\prime\prime}$ to evaluate the beam dilution effects. Under this hypothesis, the emission lines fill the beam of ALMA antennas, and the LTE-modelled spectra are generated considering the 1.3$^{\prime\prime}$ beam. Previously, \citet{cou18}, \citet{min20}, \citet{min23}, and several other authors used this method in LTE spectra to obtain the column density and excitation temperature of \ce{NH2CN} and other molecules. The FWHM of the LTE spectra of NH$_{2}$CN is 5.2$\pm$0.8 km s$^{-1}$. Our estimated excitation temperature shows that the detected transition lines of NH$_{2}$CN arise from the warm inner region of G31.41+0.31 because the temperature of the hot molecular cores exceeds 100 K \citep{van98}. We also search for rotational emission lines of NH$_{2}$$^{13}$CN and NH$_{2}$C$^{15}$N in the molecular spectra of G31.41+0.31, using the LTE model, but we could not detect these molecules. The upper limit column densities of NH$_{2}$$^{13}$CN and NH$_{2}$C$^{15}$N are $\leq$3.2$\times$10$^{13}$ cm$^{-2}$ and $\leq$5.8$\times$10$^{13}$ cm$^{-2}$.

To determine the fractional abundance of NH$_{2}$CN, we use the column density of NH$_{2}$CN inside the 1.3$^{\prime\prime}$ beam, which is divided by the column density of H$_{2}$. The abundance of NH$_{2}$CN with respect to H$_{2}$ towards G31.41+0.31 is (7.21$\pm$1.46)$\times$10$^{-10}$, whereas the column density of H$_{2}$ towards G31.41+0.31 is (1.0$\pm$0.2)$\times$10$^{25}$ cm$^{-2}$ \citep{min20}. The column density ratio of NH$_{2}$CN and NH$_{2}$CHO towards G31.41+0.31 is 0.13$\pm$0.02, where the column density of NH$_{2}$CHO in G31.41+0.31 is (5.4$\pm$1.1)$\times$10$^{16}$ cm$^{-2}$ \citep{col21}. We derived the NH$_{2}$CN/NH$_{2}$CHO ratio because NH$_{2}$CN and NH$_{2}$CHO share NH$_{2}$ as a common precursor.

\begin{table*}
	\centering
	\caption{Spectral line properties of NH$_{2}$CN towards the G31.41+0.31.}
	\begin{adjustbox}{width=1.0\textwidth}
		\begin{tabular}{ccccccccccccccccc}
			\hline 
Observed frequency &Transition & $E_{u}$ & $A_{ij}$ &S$\mu^{2}$ &FWHM&Peak intensity& V$_{LSR}$ &Optical depth& Remark\\
			
			(GHz) &(${\rm J^{'}_{K_a^{'}K_c^{'}}}$--${\rm J^{''}_{K_a^{''}K_c^{''}}}, v$) &(K) &(s$^{-1}$)&(Debye$^{2}$)&( km s$^{-1}$)&(K)&(km s$^{-1}$) & ($\tau$) & \\
			\hline
99.747&5(2,4)--4(2,3) $v$ = 1&142.19&7.87$\times$10$^{-5}$&224.84&5.25$\pm$0.32&4.19&97.02 &3.01$\times$10$^{-2}$&Non blended	\\
			
99.750&5(2,3)--4(2,2) $v$ = 1&142.19&7.87$\times$10$^{-5}$&224.83&5.24$\pm$0.48&4.25&97.01 &3.02$\times$10$^{-2}$&Non blended	\\
			
~99.825$^{*}$&5(4,1)--4(4,0) $v$ = 1&311.46&3.45$\times$10$^{-5}$&~~98.39&5.25$\pm$0.42&5.87&97.02 &6.68$\times$10$^{-3}$&Non blended	\\
			
~99.846$^{*}$&5(3,3)--4(3,2) $v$ = 1&212.77&6.14$\times$10$^{-5}$&~~58.26&5.23$\pm$0.29&1.95&97.01  &6.99$\times$10$^{-3}$&Non blended	\\
			
99.892&5(0,5)--4(0,4) $v$ = 1&~~85.70&9.63$\times$10$^{-5}$&273.87&5.25$\pm$0.65&3.87  &97.05 &2.65$\times$10$^{-2}$&Non blended	\\
			
~99.926$^{*}$&5(4,1)--4(4,0) $v$ = 0&245.86&3.56$\times$10$^{-5}$&~~33.75&--&--  &97.03 &2.77$\times$10$^{-3}$&Blended with CH$_{3}$CDO	\\
			
99.953&5(2,4)--4(2,3) $v$ = 0&~~72.36&8.30$\times$10$^{-5}$&~~78.57&5.24$\pm$0.98&1.76 &97.01 &1.29$\times$10$^{-2}$&Non blended\\
			
99.956&5(2,3)--4(2,2) $v$ = 0&~~72.36&8.30$\times$10$^{-5}$&~~78.57& --& -- &97.03 &1.23$\times$10$^{-2}$&Blended with \ce{CH3CHO}	\\
			
99.972&5(0,5)--4(0,4) $v$ = 0&~~14.39&9.90$\times$10$^{-5}$&~~93.61&--&--  &97.06 &1.67$\times$10$^{-2}$&Blended with \ce{C2H5OH}	\\
			
~100.071$^{*}$&5(3,3)--4(3,2) $v$ = 0&144.74&6.44$\times$10$^{-5}$&182.33&--&--  &97.02 &1.94$\times$10$^{-2}$&Blended with HC$_{3}$N	\\
			
			\hline
		\end{tabular}	
	\end{adjustbox}
	\label{tab:MOLECULAR DATA}\\
	{{*}}--The following transitions of NH$_{2}$CN contain double with frequency difference $\leq$100 kHz. The second transition is not shown.\\
\end{table*}

\begin{figure*}
	\centering
	\includegraphics[width=1.0\textwidth]{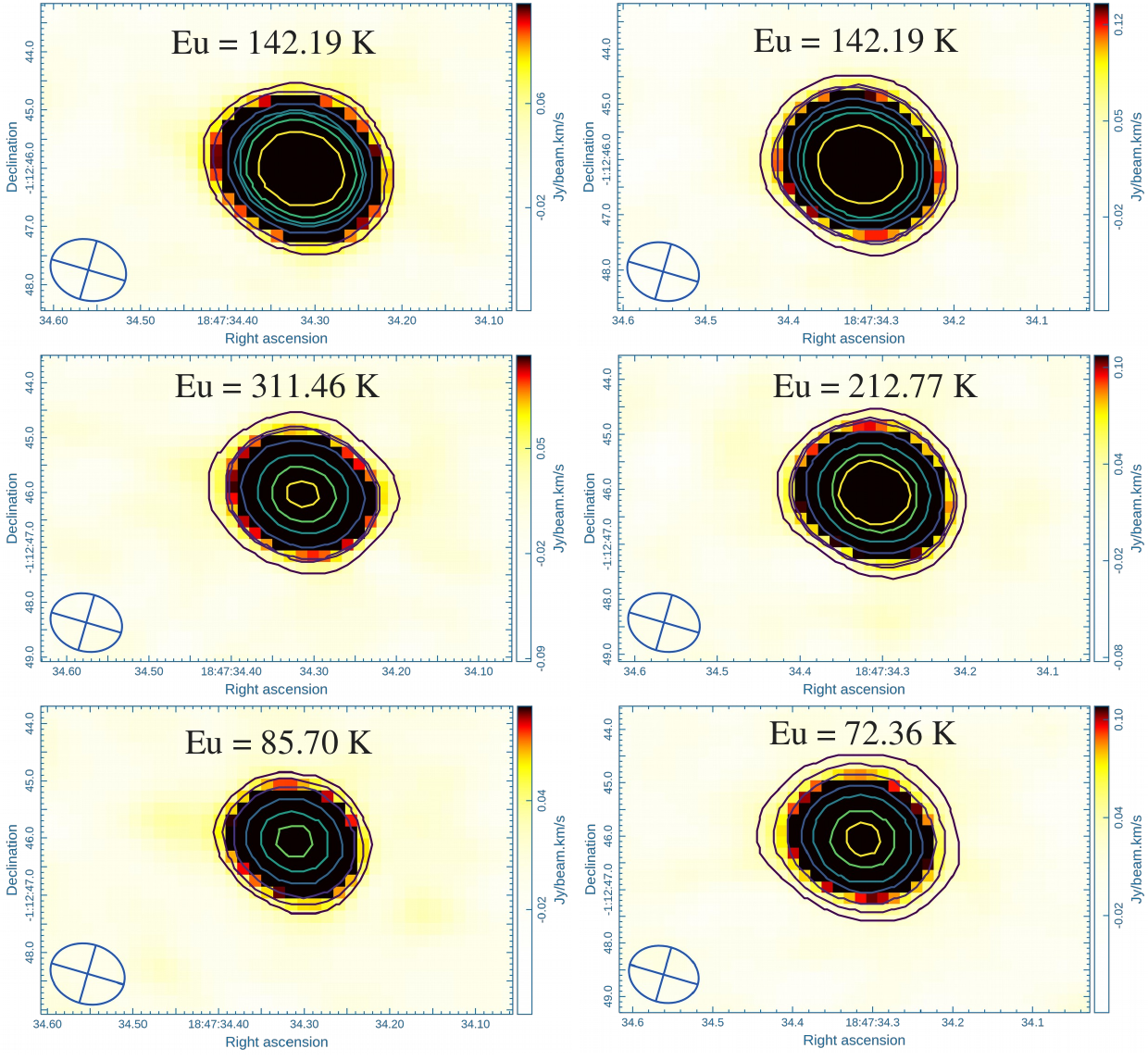}
	\caption{Integrated emission maps of NH$_{2}$CN towards G31.41+0.31. The contours indicate the 3.02 mm continuum emission map of G31.41+0.31, which is overlaid on the integrated emission maps of NH$_{2}$CN. The contour levels begin at 3$\sigma$ and increase by the factor of $\surd$2. Blue ellipses indicate the synthesized beams of the integrated emission maps.}
	
	\label{fig:emi}
\end{figure*}

\subsection{3.3 Spatial distribution of NH$_{2}$CN towards G31.41+0.31}
We created integrated emission maps of \ce{NH2CN} towards G31.41+0.31 using the CASA task {\tt IMMOMENTS}. We used six non-contaminated lines of \ce{NH2CN} to produce the emission maps. In task {\tt IMMOMENTS}, we applied the channel ranges of the spectral data cubes where the rotational emission lines of \ce{NH2CN} were detected. We constructed integrated emission maps of NH$_{2}$CN at the frequencies of 99.747 GHz, 99.750 GHz, 99.825 GHz, 99.846 GHz, 99.892 GHz, and 99.953 GHz towards G31.41+0.31, as presented in Figure~\ref{fig:emi}. We overlay the 3.02 mm continuum emission map of G31.41+0.31 over the emission maps of NH$_{2}$CN. The integrated emission maps clearly indicate that the rotational emission lines of NH$_{2}$CN arise from the warm inner region of the G31.41+0.31. We used task {\tt IMFIT} to fit the 2D Gaussian over the integrated emission maps of NH$_{2}$CN to estimate the size of the emitting regions of NH$_{2}$CN. The following equation was used to calculate the emission regions of NH$_{2}$CN,

\begin{equation} 
\theta_{S}=\sqrt{\theta^2_{50}-\theta^2_{\text{beam}}} 
\end{equation}
In the above equation, $\theta_{50} = 2\sqrt{A/\pi}$ indicates the diameter of the circle whose area ($A$) is surrounded by the $50\%$ line peak of NH$_{2}$CN, and $\theta_{\text{beam}}$ indicates the half-power width of the synthesized beam \citep{riv17}. The estimated emission regions of NH$_{2}$CN at different frequencies are presented in Table~\ref{tab:emitting region}. The emitting regions of NH$_{2}$CN vary between 1.28$^{\prime\prime}$ and 1.32$^{\prime\prime}$. Our estimated emitting regions of \ce{NH2CN} are nearly similar to the emitting regions of \ce{CH3COOH} and \ce{CH2OHCHO} in G31.41+0.31, which are estimated by \citet{min20}. We observe that the emitting regions of NH$_{2}$CN are slightly larger than, and comparable to, the beam sizes of the emission maps. This result indicates that the detected non-blended transitions of NH$_{2}$CN are not spatially resolved or marginally resolved towards G31.41+0.31. Therefore, understanding the morphology of NH$_{2}$CN from the spatial distribution maps towards G31.41+0.31 is not possible. Higher angular and spatial resolution observations are needed to determine the spatial distribution of NH$_{2}$CN towards the hot molecular core G31.41+0.31. Previously, \citet{bel18} observed the intensity dip in the integrated emission map of \ce{CH3CN} towards the center of G31.41+0.31. We did not observe that intensity dip in the emission map of \ce{NH2CN} towards the G31.41+0.31 probably due to coarse resolution.

\begin{table*}
	\caption{Emitting regions of NH$_{2}$CN towards G31.41+0.31.}
	\begin{adjustbox}{width=0.7\textwidth}
		\begin{tabular}{ccccccccccccccccc}
			\hline 
			Observed frequency&Transition&Deconvolved size&Velocity ranges\\
			(GHz)            &(${\rm J^{'}_{K_a^{'}K_c^{'}}}$--${\rm J^{''}_{K_a^{''}K_c^{''}}}, v$)             &($^{\prime\prime}$)&(km s$^{-1}$) \\
			\hline
			99.747&5(2,4)--4(2,3) $v$ = 1&1.28&93.71 to 99.31\\
			99.750&5(2,3)--4(2,2) $v$ = 1&1.30&93.25 to 99.63\\
			~99.825$^{*}$&5(3,3)--4(3,2) $v$ = 1&1.29&92.26 to 99.75\\
			~99.846$^{*}$&5(0,5)--4(0,4) $v$ = 1&1.31&92.92 to 99.39\\
			99.892&5(2,4)--4(2,3) $v$ = 0&1.32&92.09 to 99.85\\
			99.953&5(2,3)--4(2,2) $v$ = 0&1.29&92.23 to 99.56\\	
			\hline
		\end{tabular}	
	\end{adjustbox}
	\label{tab:emitting region}\\
	*--The following transitions of NH$_{2}$CN contain double with frequency difference $\leq$100 kHz. The second transition is not shown.\\
\end{table*} 

\section{4. Discussion}
\label{dis}
\subsection{4.1 Comparison of the NH$_{2}$CN abundance between hot core G31.41+0.31 and other objects}
Here, we compare the derived fractional abundance of NH$_{2}$CN in G31.41+0.31 with other hot molecular cores and hot corino objects. Previously, the rotational emission lines of NH$_{2}$CN were detected in hot molecular cores Sgr B2 (M), Sgr B2 (N), IRAS 20126+4104, G10.47+0.03, and G358.93--0.03 MM1, with estimated fractional abundances of 3.54$\times$10$^{-12}$, 5.13$\times$10$^{-9}$, 1.2$\times$10$^{-9}$, 5.07$\times$10$^{-8}$, and (4.72$\pm$2.0)$\times$10$^{-10}$, respectively \citep{bel13, bel20, pa17, man22, man23a}. Similarly, the fractional abundances of NH$_{2}$CN towards the hot corino objects NGC 1333 IRAS4A2, NGC 1333 IRAS2A, and IRAS 16293--2422 B were 1.7$\times$10$^{-10}$,  5.0$\times$10$^{-11}$, and 2.0$\times$10$^{-10}$, respectively \citep{cou18, bel20}. We estimate that the fractional abundance of NH$_{2}$CN towards the G31.41+0.31 is (7.21$\pm$1.46)$\times$10$^{-10}$. This comparison indicates that the abundance of NH$_{2}$CN in the hot core G31.41+0.31 is similar to that of the hot core G358.93--0.03 MM1 and the hot corinos NGC 1333 IRAS4A2 and IRAS 16293--2422 B. The abundance of NH$_{2}$CN in G31.41+0.31 is nearly one order of magnitude higher than NGC 1333 IRAS2A and two orders of magnitude higher than Sgr B2 (M). We also observed that the abundance of NH$_{2}$CN towards G31.41+0.31 is nearly one order of magnitude lower than that of the hot molecular cores IRAS 20126+4104 and Sgr B2 (N) and two orders of magnitude lower than that of another hot molecular core, G10.47+0.03.

\subsection{4.2 Previous chemical modelling of NH$_{2}$CN in the hot molecular cores}
Recently, \citet{zh23} used a three-phase (gas + dust surface + icy mantle) warm-up chemical model using the gas grain chemistry code {\tt NAUTILUS} to understand the abundance and possible formation mechanism of NH$_{2}$CN in hot molecular cores. \citet{zh23} assumed the free-fall collapse of a cloud (phase I), followed by a warm-up phase (phase II). In the first phase, the gas density continued to increase from 3$\times$10$^{3}$ cm$^{-3}$ to 1.6$\times$10$^{7}$ cm$^{-3}$, and a constant temperature of 10 K was maintained. In the second phase (warm-up stage), the gas density remained constant at 1.6$\times$10$^{7}$ cm$^{-3}$ but the temperature increased over time, from 10 to 200 K. During chemical modelling, \citet{zh23} used the reaction between NCN and H$_{2}$ (NCN + H$_{2}$ $\rightarrow$ NH$_{2}$CN) to produce NH$_{2}$CN on the grain surface. After chemical modelling, \citet{zh23} estimated the modelled abundance of NH$_{2}$CN towards hot molecular cores using the reaction between NCN and H$_{2}$ to be $\sim$2.8$\times$10$^{-8}$ (see Figures 3 and 4 in \citet{zh23}). The modelled abundance of NH$_{2}$CN matches with the observed abundance of NH$_{2}$CN in G10.47+0.31. The modelled abundance did not match that of other hot molecular cores, including G31.41+0.31. The chemical model of \citet{zh23} is not suitable for G31.41+0.31 because the density of G31.41+0.31 is $\sim$1$\times$10$^{8}$ cm$^{-3}$ \citep{cer10, min20}. The chemical model of \citet{zh23} is also not believable because the parent molecule NCN was not detected in the ISM until today. This means that new chemical modelling is needed to understand the modelled abundance and formation mechanism of NH$_{2}$CN towards the G31.41+0.31 and other hot molecular cores.

\begin{figure*}
	\centering
	\includegraphics[width=1.0\textwidth]{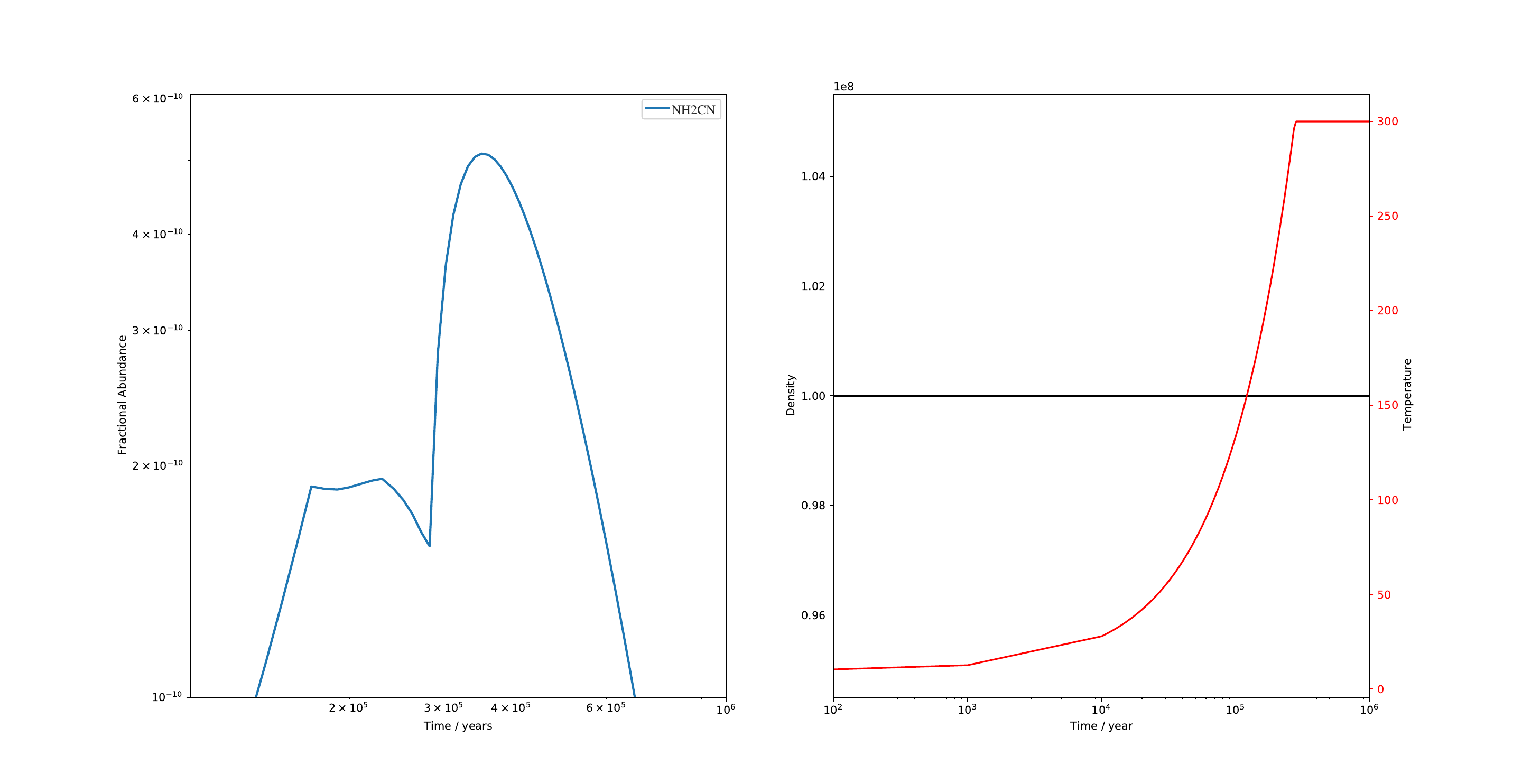}
	\caption{Two-phase warm-up chemical model abundance of NH$_{2}$CN with respect to time on the surface of grains in hot molecular cores. After warm-up phase II, the final gas density is 1$\times$10$^{8}$ cm$^{-3}$. The red line indicates the temperature profile.}
	\label{fig:model}
\end{figure*}

\subsection{4.3 New chemical modelling of NH$_{2}$CN in the hot molecular cores}
We computed a two-phase warm-up chemical model to determine the abundance and possible formation mechanisms of \ce{NH2CN}. For chemical modelling, we use the time-dependent gas-grain chemical code UCLCHEM \citep{hol17}. UCLCHEM is a time-dependent gas-grain chemical model code that focuses on grain surface chemistry as well as gas-phase reactions towards molecular clouds, hot molecular cores, and hot corinos \citep{hol17}. This chemical model includes thermal and non-thermal desorption, freeze-out, gas phase, and grain-surface reaction networks. This chemical code solves the rates of the reactions and estimates the fractional abundances of simple and complex molecules with respect to time in the grain-surface and gas-phase \citep{hol17}. In the two-phase warm-up chemical model, the free-fall collapse of the cloud (Phase I) is followed by a warm-up phase (Phase II). Our chemical model is similar to \citet{vi04}. In the first phase (Phase I), the gas density continuously rises from $n_{H}$ = 300 cm$^{-3}$ to 1$\times$10$^{8}$ cm$^{-3}$, and the gas and grain temperatures remain constant at 10 K. During the chemical modelling, the cosmic ray ionization rate was 1.3$\times$10$^{-17}$ s$^{-1}$ and the initial visual extinction ($A_{V}$) was 2. During this time, the molecules and atoms accreted on the surface of the grains at an accretion rate of 10$^{-5}$ \textup{M}$_{\odot}$ yr$^{-1}$ \citep{vi04}. The accretion rate depended on the gas density of the hot cores. In our chemical model, we assumed the sticking probability to be unity, which means that all incoming hydrogen atoms stick to the grain surface if they get an empty site. During this time, the chemical species may rapidly react or hydrogenate with other species on the grain surface. In our model, the abundances of oxygen (O), carbon (C), nitrogen (N), and helium (He) corresponding to the solar values were obtained from \citet{as09}. Other atomic compounds, such as silicon (Si), sulphur (S), chlorine (Cl), magnesium (Mg), phosphorus (P), and fluorine (F) are depleted by factors of 100. In the second phase (phase II), the temperature raised from 10 to 300 K with a constant gas density of 1$\times$10$^{8}$ cm$^{-3}$. Phase II is the warm-up stage of the hot core, and in this stage the temperature increases according to the formula $T = T_0 + (T_{max}-T_0)(\Delta t/t_h)^n$ \citep{gar06, gar08, gar13}. In this stage, the dust and gas temperatures are assumed to be well coupled. The following trends are described in \citet{cou18}, \citet{vi04}, and \citet{gar13}. In phase II, the molecules no longer freeze, and the frozen molecules on the dust grains are left in the gas phase by both non-thermal and thermal desorption mechanisms. Thermal evaporation analysis in our chemical model is described in detail in \citet{vi04} and \citet{hol17}. We also include volcano desorption, co-desorption with \ce{H2O}, and monomolecular desorption in our model, as described in detail in \citet{col04}.

For chemical modelling, we added the following reactions for the production of NH$_{2}$CN to the UCLCHEM chemical network. \\\\
\ce{NH2} + CN $\rightarrow$ NH$_{2}$CN ~~~~~~(1)\\\\
and\\\\
NH$_{2}$CNH$^{+}$ + e$^{-}$ $\rightarrow$ NH$_{2}$CN + H~~~~~~(2)\\\\
Reaction 1 indicates that the neutral-neutral reaction between NH$_{2}$ and CN produces NH$_{2}$CN on the grain surface \citep{gar06, gar13, cou18}. Earlier \citet{man23a}, \citet{cou18}, \citet{gar13}, \citet{gar06}, and \citet{gar08} claimed reaction 1 is the most efficient to produce \ce{NH2CN} in the grain surface of hot molecular cores and hot corinos. Previously, \citet{man23a} and \citet{cou18} showed reaction 1 is responsible for the production of \ce{NH2CN} towards the hot core G358.93--0.03 MM1 and the hot corino IRAS 16293 B. We claim that reaction 1 may be efficient for the formation of \ce{NH2CN} towards G31.41+0.31 because the parent molecule \ce{NH2} produces \ce{NH3} by radical recombination with hydrogen atoms and \ce{NH2} or by hydrogen abstraction from \ce{H2} molecules \citep{gar06} and previously \citet{ce92} detected the rotational emission lines of \ce{NH3} from G31.41+0.31 and other hot cores. Reaction 2 requires that the dissociative recombination of NH$_{2}$CNH$^{+}$ creates \ce{NH2CN} in the gas phase, but earlier, \citet{cou18} and \citet{gar13} showed that reaction cannot produce NH$_{2}$CN in the gas phase. The rate coefficients of reactions 1 and 2 are taken from \citet{cou18} and \citet{gar13}. The used activation energy barrier ($E_{a}$) and binding energy ($E_{b}$) for reactions 1 and 2 are 0 K and 5556 K, which are taken from \citet{cou18} and \citet{gar06}. After chemical modelling, we found that the neutral-neutral reaction between NH$_{2}$ and CN produces a sufficient amount of NH$_{2}$CN on the grain surface of hot molecular cores, but dissociative recombination of NH$_{2}$CNH$^{+}$ does not produce \ce{NH2CN} in either the gas phase or the grain surface. The computed chemical model is shown in Figure~\ref{fig:model}. The maximum modelled abundance of NH$_{2}$CN reaches 5.12$\times$10$^{-10}$ on the grain at the time of $\sim$3.8$\times$10$^{5}$ yr, which is obtained from reaction 1 in the warm-up phase. Earlier, \citet{gar22} also computed the three-phase warm-up chemical model of \ce{NH2CN} using reaction 1 with gas density 1$\times$10$^{8}$ cm$^{-3}$ and maximum temperature 400 K. After the chemical model, \citet{gar22} obtained that the abundance of \ce{NH2CN} in the warm-up conditions to be (4.1--5.0)$\times$10$^{-10}$, which is very close to our derived two-phase warm-up model abundance of \ce{NH2CN}. This indicates that reaction 1 is the most efficient pathway to the formation of \ce{NH2CN} in both two-phase and three-phase warm-up conditions.

\subsection{4.4 Comparision between observed and modelled abundance of \ce{NH2CN}}
Here, we compare our estimated abundance of \ce{NH2CN} towards G31.41+0.31 with our computed two-phase warm-up chemical model abundance of \ce{NH2CN}. Our chemical model of \ce{NH2CN} resonates with G31.41+0.31 because the gas density and temperature of G31.41+0.31 are $\sim$1$\times$10$^{8}$ cm$^{-3}$ and $\geq$150 K \citep{min20}. From the chemical modelling, we determine that the abundance of \ce{NH2CN} is 5.12$\times$10$^{-10}$. We estimate the fractional abundance of \ce{NH2CN} towards the G31.41+0.31 to be (7.21$\pm$1.46)$\times$10$^{-10}$. This result indicates that the estimated abundance of \ce{NH2CN} towards the G31.41+0.31 is nearly similar to the modelled abundance of \ce{NH2CN}. It shows \ce{NH2CN} is created towards the G31.41+0.31 via the neutral-neutral reaction between \ce{NH2} and CN on the grain surface of the hot core. Recently, \citet{man23a} determined that the fractional abundance of \ce{NH2CN} towards another hot molecular core G358.93-0.03 MM1 is (4.72$\pm$2.0)$\times$10$^{-10}$, which is similar to the modelled abundance of \ce{NH2CN}. This indicates, like G31.41+0.31, \ce{NH2CN} is created via the neutral-neutral reaction between \ce{NH2} and CN on the grain surface towards the other hot molecular core, G358.93--0.03 MM1.

\section{5. Conclusion}
\label{conclu}
In this study, we analyze the molecular spectra of the chemically rich hot molecular core G31.41+0.31, which was observed using high-resolution ALMA at band 3. For the first time, we identified the rotational emission lines of both the cyan and amide-containing molecule \ce{NH2CN} in the molecular spectra of G31.41+0.31. The estimated excitation temperature and column density of \ce{NH2CN} for G31.41+0.31, using the local thermodynamic equilibrium (LTE) model, are 250$\pm$25 K and (7.21$\pm$0.25)$\times$10$^{15}$ cm$^{-2}$, respectively. The abundance of NH$_{2}$CN towards the G31.41+0.31 with respect to H$_{2}$ is (7.21$\pm$1.46)$\times$10$^{-10}$.  The NH$_{2}$CN/NH$_{2}$CHO ratios towards the G31.41+0.31 is 0.13$\pm$0.02. We also created the integrated emission maps of \ce{NH2CN}, and we noticed that the emission lines of \ce{NH2CN} originate from the warm and inner parts of G31.41+0.31. The size of the emitting regions of \ce{NH2CN} vary between 1.28$^{\prime\prime}$ and 1.32$^{\prime\prime}$. We compare our derived fractional abundance of \ce{NH2CN} with that of other hot core and corino objects. After comparison, we find that the abundance of \ce{NH2CN} towards the G31.41+0.31 is similar to the abundance of \ce{NH2CN} towards the molecular core G358.93--0.03 MM1 and hot corinos NGC 1333 IRAS4A2 and IRAS 16293--2422 B. We also compute the two-phase warm-up chemical model of \ce{NH2CN} using the gas-grain chemical code UCLCHEM. After chemical modelling, we find that the modelled abundance of \ce{NH2CN} towards the hot molecular cores is 5.12$\times$10$^{-10}$. This result shows that the observed abundance of \ce{NH2CN} towards the G31.41+0.31 is similar to the modelled abundance of \ce{NH2CN}. This implies that NH$_{2}$CN have been created on the grain surface of G31.41+0.31 via a neutral-neutral reaction between \ce{NH2} and CN. The detection of \ce{NH2CN} towards the G31.41+0.31 indicates that grain-surface chemistry is efficient for the formation of other complex organic molecules, including other complex nitrogen-bearing molecules in this hot core. A spectral lines observation of other nitrogen-bearing molecules using the ALMA with better spectral resolution is needed towards G31.41+0.31 to understand the nitrile chemistry in this hot molecular core, which will be carried out in our follow-up study.

\section*{Author information}
\text{Corresponding Author}: Arijit Mnna\\
\text{E-mail: arijitmanna@mcconline.org.in}\\
\text{ORCID}:\\
\text{Arijit Manna: 0000-0001-9133-3465}\\
\text{Sabyasachi Pal: 0000-0003-2325-8509}\\
\text{Notes: The authors declare no competing financial interest.}

\begin{acknowledgement}
We thank the anonymous referees for their helpful comments, which improved the manuscript. A.M. acknowledges the Swami Vivekananda Merit-cum-Means Scholarship (SVMCM), Government of West Bengal, India, for financial support for this research. This paper makes use of the following ALMA data: ADS /JAO.ALMA\#2017.1.00501.S. ALMA is a partnership of ESO (representing its member states), NSF (USA), and NINS (Japan), together with NRC (Canada), MOST and ASIAA (Taiwan), and KASI (Republic of Korea), in co-operation with the Republic of Chile. The Joint ALMA Observatory is operated by ESO, AUI/NRAO, and NAOJ.

\end{acknowledgement}



\section{Graphical Abstract}
\begin{figure}
	\centering
	\includegraphics[width=1.0\textwidth]{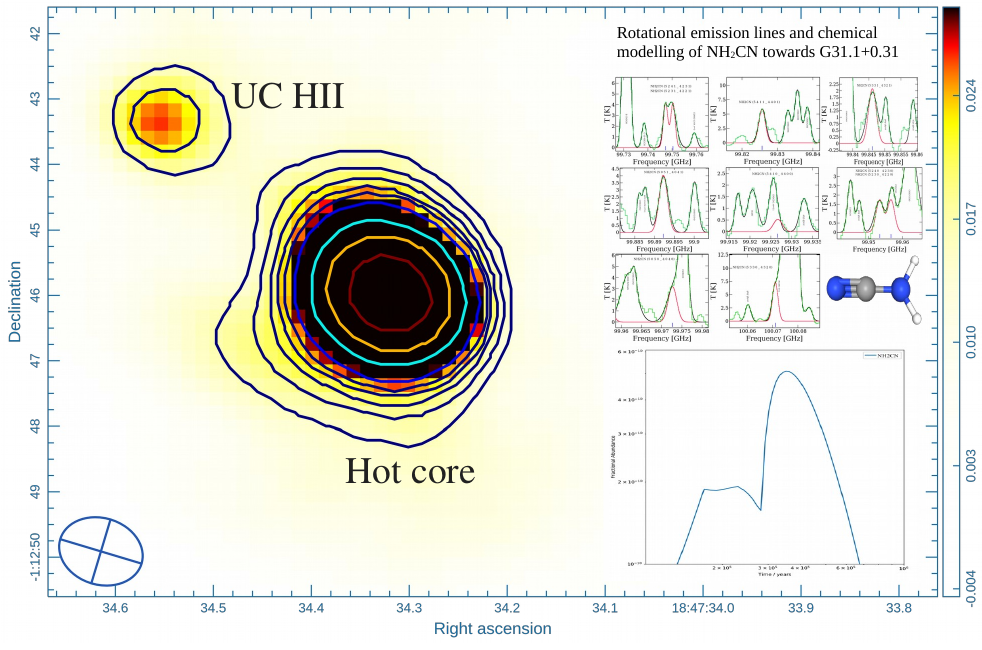}
\end{figure}
\end{document}